\documentclass[trackchanges,twocolumn]{aastex701}

\newcommand{\Sg}{Sgr~A*}

\defcitealias{Neilsen2013}{N13}
\usepackage{float}
\usepackage{amsmath}
\usepackage{rotating}

\begin{document}

\title{The Demographics of Sagittarius A* X-ray Flares over 25 Years with Chandra}

\author[0009-0004-8539-3516]{Zach Sumners}
\affiliation{McGill University, Montreal QC H3A 0G4, Canada}%
\affiliation{Trottier Space Institute, 3550 Rue University, Montréal, Québec, H3A 2A7, Canada}
\email[show]{ronald.sumners@mail.mcgill.ca}

\author[0000-0001-8921-3624]{Nicole M. Ford}
\affiliation{McGill University, Montreal QC H3A 0G4, Canada}%
\affiliation{Trottier Space Institute, 3550 Rue University, Montréal, Québec, H3A 2A7, Canada}
\email[]{nicole.ford@mail.mcgill.ca}

\author[0000-0001-6803-2138]{Daryl Haggard}
\affiliation{McGill University, Montreal QC H3A 0G4, Canada}%
\affiliation{Trottier Space Institute, 3550 Rue University, Montréal, Québec, H3A 2A7, Canada}
\email{daryl.haggard@mcgill.ca}

\author[orcid=0000-0003-3503-3446,sname=Michail, gname=Joseph]{Joseph M. Michail}
\altaffiliation{NSF Astronomy \& Astrophysics Postdoctoral Fellow}
\affiliation{Center for Astrophysics $|$ Harvard \& Smithsonian, 60 Garden Street, Cambridge, MA 02138, USA}
\email{joseph.michail@cfa.harvard.edu}  

\author[0000-0002-8247-786X]{Joey Neilsen}
\affiliation{Department of Physics, Villanova University, 800 Lancaster Avenue, Villanova, PA 19085, USA}
\email{joseph.neilsen@villanova.edu}

\author[0000-0001-6923-1315]{Michael A. Nowak}
\affiliation{Department of Physics, Washington University in St. Louis, Campus Box 1105, One Brookings Drive, St. Louis, MO 63130-4899, USA}
\email{mnowak@physics.wustl.edu}

\author[orcid=0000-0001-9641-6550,sname='Balakrishnan']{Mayura Balakrishnan}
\affiliation{McGill University, Montreal QC H3A 0G4, Canada}%
\affiliation{Trottier Space Institute, 3550 Rue University, Montréal, Québec, H3A 2A7, Canada}
\email{mayura.balakrishnan@mcgill.ca}

\author[0000-0002-0947-569X]{Sophia S\'{a}nchez-Maes}
\affiliation{Institute for Research in Electronics and Applied Physics, University of Maryland, 8279 Paint Branch Drive, College Park, MD 20742, USA}%
\email{sophiasm@umd.edu}

\author[orcid=0000-0002-9156-2249,sname='Sebastiano']{Sebastiano D. von Fellenberg}
\altaffiliation{Feodor Lynen Fellow}
\affiliation{Canadian Institute for Theoretical Astrophysics, University of Toronto, 60 St. George Street, Toronto, ON M5S 3H8, Canada}
\affiliation{Max Planck Insitute for Radioastronomy, auf dem H{\"u}gel 69, Bonn, Germany }
\email{sfellenberg@cita.utoronto.ca}

\author[0000-0002-9895-5758]{S. P. Willner}
\affiliation{Center for Astrophysics $|$ Harvard \& Smithsonian, 60 Garden Street, Cambridge, MA 02138, USA}
\email{swillner@cfa.harvard.edu}

\author[0000-0001-9564-0876]{Sera Markoff}
\affiliation{Anton Pannekoek Institute for Astronomy, University of Amsterdam, Science Park 904, 1098 XH Amsterdam, The Netherlands}
\affiliation{Gravitation and Astroparticle Physics Amsterdam Institute, University of Amsterdam, Science Park 904, 1098 XH 195 196 Amsterdam, The Netherlands}
\affiliation{Institute of Astronomy, University of Cambridge, Madingley Road, Cambridge CB3 0HA, United Kingdom}
\email{S.B.Markoff@uva.nl}

\author[]{Howard A. Smith}
\affiliation{Center for Astrophysics $|$ Harvard \& Smithsonian, 60 Garden Street, Cambridge, MA 02138, USA}
\email{hsmith@cfa.harvard.edu}

\author[0000-0002-5599-4650]{Joseph L. Hora}
\affiliation{Center for Astrophysics $|$ Harvard \& Smithsonian, 60 Garden Street, Cambridge, MA 02138, USA}
\email{jhora@cfa.harvard.edu}


\begin{abstract}
We present the \textit{Chandra} 25-year Sagittarius A* (\Sg) X-ray flare catalog: a systematic analysis of 6.8~Ms of \Sg\ monitoring spanning the \textit{Chandra} X-ray Observatory's mission lifetime. This is the most complete \textit{Chandra} \Sg\ X-ray flare catalog to date, consisting of 100 flares with 2$-$10 keV unabsorbed luminosities ranging from $\sim$\,4$-$575\,$\times\,10^{33}$ 
erg s$^{-1}$. 18 flares are reported for the first time, including the second brightest \Sg\ flare observed by \textit{Chandra}. The expanded dataset supports previous indications of a correlation between X-ray flare hardness and luminosity. Spectral modeling corroborates this finding, showing a change in the X-ray spectral index, from $\Gamma\,\sim3$ to 2 with increasing flare brightness. Previously-established correlations between flare duration, fluence, and maximum count rate are strengthened via the greater sample size. 
These results likely reflect variations in the underlying particle distribution that produce weak and strong flares, and the
new catalog serves as a rich archive for ongoing observational and numerical investigations into the physical mechanisms responsible for producing \Sg's X-ray flares.

\end{abstract}



\section{Introduction}
Sagittarius A* (\Sg), the supermassive black hole (SMBH) at the center of the Milky Way Galaxy, is approximately 8~kpc from Earth with a mass of $\sim 4 \times 10^6$~M$_{\sun}$ \citep{Do2019, GRAVITY2022}. It exhibits rapid variability (``flares'') 
at radio \citep[e.g.,][]{Bower2015, Chen2023}, submillimeter \citep[e.g.,][]{Subroweit2017, Michail2024}, infrared \citep[IR; e.g.,][]{Genzel2003, Ghez2004, Hora2014, Do2019b, vonFellenberg2025, Michail2026}, and X-ray wavelengths \citep[e.g.,][]{Baganoff2001, Nowak2012, Neilsen2013, Haggard2019, Bouffard2019}. 
However, the specific mechanisms responsible for flares and how they are correlated across across frequencies remains unclear \citep[e.g.,][]{Fazio2018, GRAVITY2021, Boyce2022}. Simultaneous multi-wavelength campaigns have provided significant observational context to theoretically probe these physical drivers \citep[e.g.,][see \citealt{Ciurlo2025} for a review]{Eckart2006, Yusef-Zadeh2006, Dodds-Eden2009, Capellupo2017, GRAVITY2021, vonFellenberg2025}. 
This modeling has shown that flares may arise from nonthermal electrons radiating via synchrotron emission with a high energy tail and a cooling break \citep{Yuan2003, Yusef-Zadeh2006, Dodds-Eden2010, Ponti2017}, and/or synchrotron self Compton emission where synchrotron IR photons are upscattered to higher energies \citep{Markoff2001, Liu2002, Eckart2008, Witzel2021}. A more complete description of the total population of X-ray flares is key to understanding when these proposed emission mechanisms may be operating. 


At the angular resolution of the \textit{Chandra} X-ray Observatory, \Sg\ is spatially distinguishable from other objects in the Galactic Center and has a quiescent (non-flaring) 2--10~keV unabsorbed luminosity of $3 \times 10^{33}$~erg~s$^{-1}$ \citep{Baganoff2003, Nowak2012, Boyce2019}. This quiescence is disrupted by an X-ray flare approximately once every day, with peak fluxes observed up to hundreds of times the non-flaring level with typical durations of minutes to hours \citep{Neilsen2013, Ponti2015, Mossoux2017, Mossoux2020}. \textit{Chandra} observed the first X-ray flare from \Sg\ in late 2000 \citep{Baganoff2001} and since then, 174 \textit{Chandra} observations with the SMBH in the field, totaling $\sim 6.8$ Ms, have been completed. \Sg's quiescent state is modeled as a power law with photon X-ray spectral index\footnote{The X-ray spectral index is related to infrared by $\nu F_{\nu} \propto \nu^{2-\Gamma}$.} $\Gamma = 3$\ \citep{Nowak2012}, but the flaring state has been reported to follow a power law with index $\Gamma \approx 2$ \citep{Baganoff2001,Neilsen2015,Ponti2017,Zhang2017,Haggard2019}. However, low signal-to-noise ratios (SNRs) from weak flares and persistent instrumental effects 
have affected how well flare parameters can be measured \citep{Neilsen2013}. 

\textit{Chandra} has collected 996~ks of new \Sg\ observations since the last census by \citet{Mossoux2017} and \citet{Mossoux2020}, where they cataloged 89 \textit{Chandra} flares. 
In this study, we present new flares from this updated dataset, and also systematically reanalyze all \textit{Chandra} \Sg\ observations with updated instrument calibration files, barycenter corrections, and modeling of instrumental systematics to create the most complete catalog of \Sg\ flares to date. All the flares we report for the first time in this study are from the 996~ks of new data. We do not detect 7 low-level flares that \citet{Mossoux2017} and \citet{Mossoux2020} reported. 
This catalog enables a uniform characterization of flare properties over the lifetime of \textit{Chandra} to highlight demographic trends that point to the origin of flares, and assist future multi-wavelength temporal correlation studies. 

This paper is organized as follows.   Section~\ref{sec:data_analysis} outlines the data reduction pipeline. Section~\ref{sec:flare_demographics} presents demographic characteristics among the population of X-ray flares. Section~\ref{sec:spectral_properties} analyzes the spectral properties of brighter flares, and Section \ref{sec:discussion} discusses the observed trends in the context of flare production. 

\section{Data Analysis}
\label{sec:data_analysis}

\subsection{Data Products}
We identify 175 observations in the \textit{Chandra} source catalog \citep{Evans2024} pointed within 10\farcs0 of \Sg, captured between 1999 September 21 and 2025 August 11.
These data were captured as part of both individual PI-led proposals and larger programs such as the X-Ray Visionary Project (XVP\footnote{\url{https://www.sgra-star.com}}; a 3~Ms campaign from 2012 using the High Energy Transmission Grating; HETG), GRAVITY campaigns \citep[e.g.,][]{GRAVITY2021} and multi-wavelength observations collected in parallel with Event Horizon Telescope campaigns \citep[EHT; e.g.,][]{EHT2022b}. The archival data are spaced non-uniformly in time and span 10 observing modes, 
each with a unique combination of CCD chips, gratings, frame times and, if applicable, sub-array setups \citep{Weisskopf2002, Wilkes2024}. They broadly exist in three classes: ACIS-I with no grating, ACIS-S with the HETG, and ACIS-S with no grating. Our dataset constitutes 1.0~Ms, 2.8~Ms, and 3.0~Ms of each mode, respectively. Photons in grating observations are dispersed into spectral orders, and require additional data processing. We omit from our study three ACIS-S grating observations (ID 15040, 15651, and 15654) where the grating arms are contaminated by an outburst from magnetar SGR J1745-2900 \citep{Mori2013} and one observation of the magnetar with the \textit{Chandra} High Resolution Camera (ID 14701). Magnetar contributions to \Sg's flux in observations without the HETG are discussed in Section \ref{subsec:magnetarcorrection}.



\begin{figure*}
     \centering
     \includegraphics[width=\textwidth]{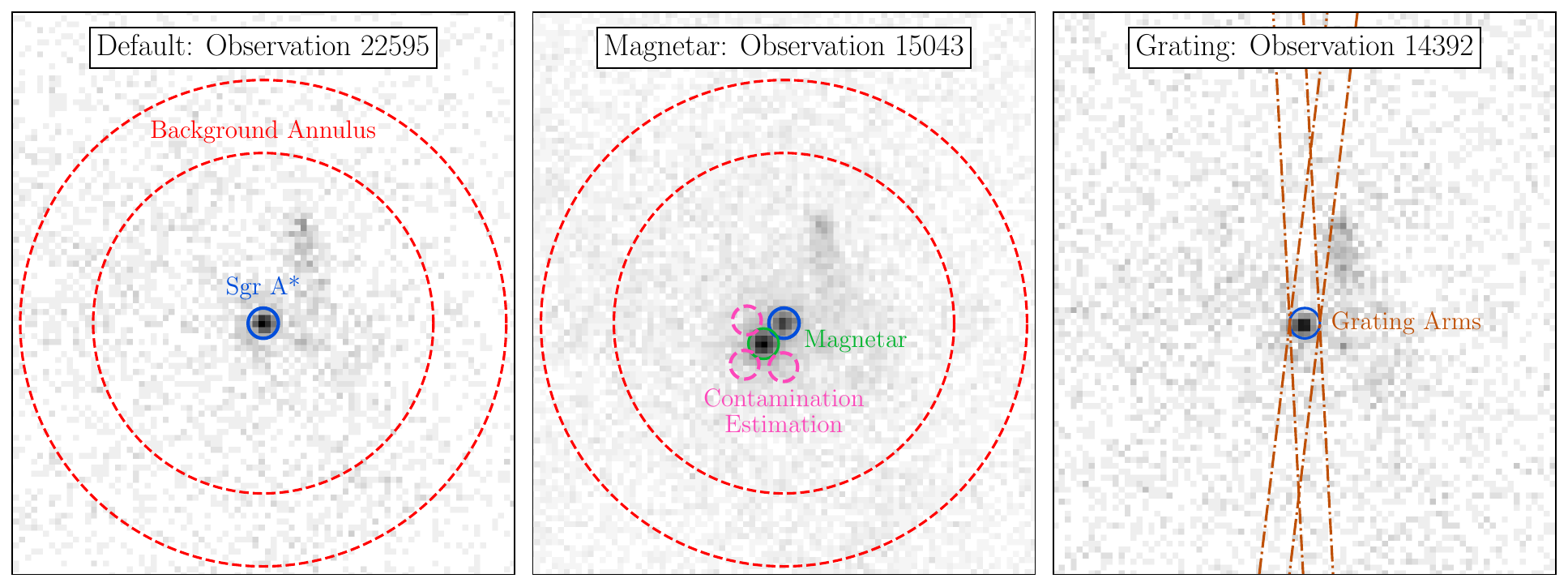}
     \caption{Three types of extraction regions are used to generate \Sg\ light curves: default, magnetar contaminated, and grating observation, respectively. \textit{Left}: The default extraction; the blue circle denotes a $1\farcs25$ radius aperture centered on \Sg\ collecting source photons, and the red dashed annulus is the background region. \textit{Middle}: Observations captured during the outburst of magnetar SGR J1745-2900 require a specialized extraction scheme to correct for flux contamination. The blue and red apertures remain as \Sg\ and the background. The central green circle marks a $1\farcs30$ radius region centered on the magnetar. The dashed pink circles are $1\farcs25$ radius magnetar contamination estimator regions, placed at the same separation from the magnetar as \Sg\ , but at opposite position angles. \textit{Right}: Grating observations disperse photons, offering better energy resolution. Zeroth-order photons are concentrated in the blue circle at the center, as in the left panel. The dispersed $\pm 1$ order photons form an X-shaped pattern extending diagonally from the center, outlined by the dash-dotted brown lines. We do not subtract a background for grating observations.}
     \label{fig:regions}
\end{figure*}

\subsection{Data Processing Pipeline}
\label{sec:processing}
The goal of this study is a comprehensive analysis of \Sg\ flares during the first $\sim 25$ years of \textit{Chandra} operations. 
We create a \textit{Chandra} data processing pipeline using the \texttt{CIAO v4.16} software library \citep[\texttt{CALDB 4.11.5};][]{Fruscione2006} and custom data corrections to perform systematic and uniform data processing to the extent that the observing setup allows. First, all data are reprocessed with the latest calibration files using the \textsc{chandra\_repro} method. We then barycenter correct each reprocessed event file with \textsc{axbary}. To flexibly locate \Sg\ given the unique pointing of each observation, \textsc{wavdetect} then identifies all point sources in the image and collects their world coordinate system (WCS) locations. We correct the astrometry of these source positions with \textsc{wcs\_match} and \textsc{wcs\_update}, and assume the corrected centroid closest to the radio coordinates of \Sg\ \citep[$17^{\rm{h}}45^{\rm{m}}40\fs0409$, -29$^{\circ}$00\arcmin28\farcs118;][]{Reid2004} is our object of interest. If \textsc{wavdetect} does not center a source within 2\arcsec\ of \Sg's coordinates (potentially due to confusion with the magnetar), we manually place a source region and visually center it on \Sg. The position of source extraction regions is visually verified for all observations and refined if required. The average positional uncertainty of \Sg\ after WCS correction is $\sim0\farcs09$ in right ascension and $0\farcs12$ in declination. This is much less than one pixel in \textit{Chandra} ACIS's field of view of $0\farcs492$ per pix \citep{Garmire2003}.

\subsubsection{Light Curve Extraction}
\label{sec:extraction}
Following \citet[][herein called \citetalias{Neilsen2013}]{Neilsen2013}, light curves are binned to 300~sec intervals in the 2$-$8~keV energy range, and are extracted from three types of regions detailed in Figure \ref{fig:regions}.

In non-grating observations when the magnetar is not in outburst, we use a 1\farcs25 radius circular region and a background annulus of inner radius 14\arcsec~and outer radius 20\arcsec~ (left panel of Figure \ref{fig:regions}). They are both centered on \Sg\ .

We follow the work of \citet{Nowak2012}, \citetalias{Neilsen2013}, and \citet{Corrales2020} for observations with the HETG, and only focus on the zeroth- and first-order. Zeroth-order photons are not dispersed and fall in the center of the field. We extract them with the same 1\farcs25 circular region as before. First-order photons are dispersed along the grating arms, and are extracted with two 2\farcs5 wide rectangular boxes centered at the zeroth-order position of \Sg, rotated in the direction of each grating arm. The right panel of Figure \ref{fig:regions} shows this setup. The background of these regions is sufficiently low such that we do not need a background subtraction region \citep[][\citetalias{Neilsen2013}]{Nowak2012}. We also use the \textsc{tg\_create\_mask} method to record the order of each photon. 

\subsubsection{Magnetar Correction}
\label{subsec:magnetarcorrection}

SGR J1745$-$2900 is a magnetar $\sim 3\farcs0$ from \Sg\ that went into outburst in 2013 \citep{Rea2013}. \Sg\ light curves between 2013 May 12 and 2016 July 12 are affected by its emission, and require special ``contamination treatment''. We follow the methodology outlined in \citet{Haggard2019} and \citet{Bouffard2019} to estimate its contribution in these cases and subsequently correct \Sg\ light curves. While the magnetar continues to be detected after 2016 July, its flux declines to a negligible contamination level at the position of \Sg\ \citep{Zelati2017}. 

We employ additional extraction regions visualized in the middle panel of Figure \ref{fig:regions} for this treatment, where a 1\farcs3 radius circular region is centered on the magnetar, and three additional 1\farcs25 radius circular regions are drawn equally separated from the magnetar, but on opposite sides. The \Sg\ and background regions remain as described in Section \ref{sec:extraction}. 
We assume the flux measured in the three 1\farcs25 ``estimator'' regions arise exclusively from the magnetar and the background, allowing us to estimate how many photons the magnetar produces at an image separation equivalent to that of \Sg. Light curves are extracted from the estimator regions, and the average of the three is subtracted from \Sg's light curve to generate a magnetar-corrected count rate, 
as in Equation 1 of \citet{Bouffard2019}.



\subsubsection{Light Curve Pileup Correction}
\label{sec:pileup}
Pileup is an instrumental effect where the number and energy of detected photons is misreported when multiple fall on the same CCD pixel during a single frame \citep{Davis2001}. The degree of pileup scales nonlinearly with incident (true) flux, and is impacted by the frame time, detector setup, and location on the CCD. Various methods have been proposed to correct for pileup in \textit{Chandra}  observations of \Sg\ (\citealt{Nowak2012}, \citetalias{Neilsen2013}, \citealt{Ponti2015,Bouffard2019}), but each is limited to a certain instrumental configuration and/or does not capture the temporal changes of the detectors over the 25 years of observing. 

In this analysis, we employ \texttt{MARX} \citep{Davis2012}, a \textit{Chandra} raytracing simulator, to estimate how pileup affects 
light curves across multiple years and observing modes. For every real \textit{Chandra} \Sg\ observation, we generate a series of simulated \texttt{MARX} measurements with a range of assumed incident fluxes. The simulations are set to match the instrumental configuration of the real observation (frame time, source type, pointing parameters, dither patterns). We compare the simulated incident count rate (no pileup effects) with the observed rate (pileup present) by toggling \textsc{marxpileup}. We assume a grade mitigation factor (how the grade assigned to each photon changes due to pileup), $\alpha_{\text{pileup}}$ = 0.5 when pileup is present \citep{Haggard2019}. Varying $\alpha_{\text{pileup}}$ does not significantly change the correction. Examples showing the resulting pileup strength (difference between incident and observed flux) as a function of observed count rate are shown in Appendix \ref{appx:pileup}. Each measured \textit{Chandra} rate is then corrected according to the \texttt{MARX} results to estimate the true count rate. In gratings observations, the zeroth- and first-orders are simulated separately and subsequently summed, though dispersion along the grating arms often leads to negligible pileup correction for the first order. 
%


\subsubsection{Flux Calibration}
\label{sec:flux_calibration}
The spectral fitting in Section \ref{sec:spectral_properties} enables flux calibration (conversion between instrumental counts and physical units) via \textsc{cflux} in \texttt{XSPEC} \citep{Arnaud1999}. 
Averaging over 37 fits, non grating observations have a conversion of 0.01188 ct/$10^{34}$~erg, and grating observations have an average calibration of 0.008545 ct/$10^{34}$~erg in the unabsorbed $2-10$~keV band (see also \citealt{Bouffard2019}). No significant difference is found between the conversion factors of nongrating ACIS-S and ACIS-I observations. As in Section \ref{sec:spectral_properties}, we only fit the zeroth order of grating data. These calibrations are applied to the count rates and fluence of their respective light curves to estimate the luminosity and energy of flares (see Table \ref{tab:tablehead}). We assume a distance of 8178~pc \citep{GRAVITY2019}.

\subsubsection{Bayesian Blocks Detection}
Bayesian Blocks is a coarse characterization algorithm that calculates a ``fitness factor'' to detect statistically significant change points in time series \citep{Scargle2013}. Studies such as \citet{Nowak2012}, \citetalias{Neilsen2013}, \citet{Ponti2015}, \citet{Mossoux2017}, \citet{Bouffard2019}, and \citet{Mossoux2020} have applied this technique to detect \Sg\ flares in \textit{Chandra} observations, and we follow their work with an implementation adapted by \citet{Williams2017} to find and characterize flares in this updated \textit{Chandra} dataset. Following \citet{Scargle2013}, \citetalias{Neilsen2013}, \citet{Ponti2015}, \citet{Mossoux2017}, and \citet{Bouffard2019}, we set the change point false alarm probability to $p_0 = 0.05$. This gives a false flare detection rate of $p_0^2 = 0.25\%$ because flares have two change points. We use bootstrap simulations created by \cite{Williams2017} to estimate errors in block heights. \citet{Scargle2013} claimed that bootstrapping the change points themselves has little effect on the diversity of blocks, so we do not calculate uncertainties in the flare start and stop times, and therefore we do not have errors on duration. Blocks are generated at all statistically significant heights, including at quiescence. Therefore, we define that a block must be 3$\sigma$ above quiescence and at least 1 time bin (300~sec) in length to be considered a flare. The lower limits of our count rates and durations are therefore subject to observational biases (see \citealt{Neilsen2015} for a discussion of faint flare biases). Flare durations are measured from the beginning to end of a single flaring block or consecutive sequence of flaring blocks. Additional discussion of Bayesian Blocks and alternative flare detection methods can be found in \citet{Nowak2012}, \citetalias{Neilsen2013}, and \citet{Ponti2015}.

\begin{deluxetable}{lccccccccccc} 
\tabletypesize{\footnotesize}
\setlength{\tabcolsep}{1.9pt}
\caption{The first 5 rows of the \textit{Chandra} 25-year Sgr A* X-ray flare catalog, in order of flare start time. The full table is available in a machine readable format. The \textit{Chandra} observation ID, flare number (for multi-flare observations), shape, time, duration, mean rate, max rate, fluence, flux, luminosity, and hardness ratio (HR) of each flare are listed.}
\label{tab:tablehead}
\tablehead{
    \colhead{ID} & \colhead{Flare} & \colhead{Shape} & \colhead{Start} & \colhead{End} & \colhead{Duration} & \colhead{Mean Rate} & \colhead{Max Rate} & \colhead{Fluence} & \colhead{Flux$^{(a)}$} & \colhead{Luminosity$^{(a)}$} & \colhead{HR}\\
    \colhead{} & \colhead{} & \colhead{} & \colhead{MJD} & \colhead{MJD} & \colhead{(s)} & \colhead{(ct s$^{-1}$)} & \colhead{(ct s$^{-1}$)} & \colhead{(ct)} & \colhead{($\times$10$^{-12}$erg s$^{-1}$cm$^{2}$)} & \colhead{($\times$10$^{34}$ erg s$^{-1}$)} & \colhead{}
}
\startdata
1561 & 1 & S & 51844.105\;& 51844.113 & 668 & 0.037\,$\pm$\,0.007\;& 0.042\,$\pm$\,0.011\;& 25 $\pm$ 5 & 3.33 $\pm$ 0.06 & 2.67 $\pm$ 0.05 & $1.3^{+0.3}_{-0.6}$ \\
1561\textsuperscript{\textdagger} & 2 & SS & 51844.158\;& 51844.280 & 10484 & 0.088\,$\pm$\,0.003\;& 0.414\,$\pm$\,0.023\;& 924 $\pm$ 30 & 8.72 $\pm$ 0.06 & 6.98 $\pm$ 0.05 & $1.9^{+0.1}_{-0.2}$ \\
3663\textsuperscript{\textdagger} & 1 & D & 52418.801\;& 52418.856 & 4735 & 0.023\,$\pm$\,0.002\;& 0.045\,$\pm$\,0.012\;& 106 $\pm$ 10 & 1.80 $\pm$ 0.05 & 1.44 $\pm$ 0.04 & $1.6^{+0.3}_{-0.4}$ \\
3392 & 1 & S & 52420.175\;& 52420.210 & 3033 & 0.023\,$\pm$\,0.003\;& 0.044\,$\pm$\,0.012\;& 70 $\pm$ 8 & 1.92 $\pm$ 0.03 & 1.54 $\pm$ 0.03 & $2.0^{+0.4}_{-0.6}$ \\
3392\textsuperscript{\textdagger} & 2 & S & 52420.569\;& 52420.631 & 5352 & 0.012\,$\pm$\,0.002\;& 0.032\,$\pm$\,0.010\;& 66 $\pm$ 8 & 0.80 $\pm$ 0.03 & 0.64 $\pm$ 0.03 & $0.8^{+0.1}_{-0.3}$ \\
\hline
\enddata
\tablenotetext{\textsuperscript{\textdagger}}{Flagged flares; see Section \ref{sec:flagging}.}
\tablenotetext{(a)}{~~Unabsorbed $2-10$ keV band.}
\tablecomments{In the shape column, ``S'' indicates a single peaked event, ``D'' is a double peaked flare, and ``SS'' is a flare with more complex substructure.}
\end{deluxetable}

\section{Flare Demographics}
\label{sec:flare_demographics}

We search 6.8~Ms of \textit{Chandra} observations and detect a total of 100 X-ray flares in 68 of the 171 observations. \Sg\ is in a flaring state about $\sim 5\%$ of the total exposure time. We measure the median quiescent count rate of \Sg\ to be 0.005 $\pm$ 0.0005 ct s$^{-1}$ (0.005 $\pm$ 0.0005 ct s$^{-1}$ for ACIS-S with no grating, 0.005 $\pm$ 0.0006 ct s$^{-1}$ for ACIS-I, and 0.006 $\pm$ 0.0004 ct s$^{-1}$ for ACIS-S observations with the HETG) which is consistent with other studies \citep[see also][]{Nowak2012, Bouffard2019}. Count rates and derived parameters (such as fluence, flux, luminosity, and energy) are presented in Table \ref{tab:tablehead}, with all values being pileup-corrected unless otherwise specified. The brightest flare was measured on 2013 September 14 and has a maximum observed count rate of 1.05 ct s$^{-1}$ \citep[first reported in][]{Haggard2019}. Pileup correction increases this rate by 79\% to 1.88 ct s$^{-1}$. The second brightest \textit{Chandra} flare on 2019 August 19, reported for the first time here, has an observed (unpiled) count rate of 0.66 ct s$^{-1}$, which increases to 0.89 ct s$^{-1}$ (35\%) after pileup correction.


The maximum corrected flare count rate spans $0.02 - 1.88$~ct s$^{-1}$, and the mean count rate varies from $0.01 - 0.69$~ct s$^{-1}$ across the detected events.
We estimate the flux, luminosity, and energy of the flares using the mean rate and flux calibration outlined in Section \ref{sec:flux_calibration}. $2-10$~keV unabsorbed flare luminosities span $4.2\times10^{33}-5.7\times10^{35}$~erg s$^{-1}$, with an average of  4.8$\times$10$^{34}$~erg s$^{-1}$. This range of luminosities is $\sim 2 - 200$ times the quiescent rate \citep{Baganoff2001, Baganoff2003, Nowak2012}. The flare fluences (integrated counts) span $16 - 3814$~ct ($1.4 \times 10^{37} - 3.2 \times 10^{39}$~erg). Modeling the occurrence of events in the last 25 years as a Poisson distribution, we confirm a flaring rate of $1.3 \pm 0.1$~flares/day, agreeing with \citetalias{Neilsen2013}'s rate of $1.1^{+0.2}_{-0.1}$~flares/day. 
Our detections span in duration from $367 -14011$~sec (6.1~minutes $-$ 3.9~hours), but faint flares are challenging to characterize accurately, as we discuss further in Section \ref{sec:flagging} and \ref{sec:parameter_distributions}, so this range may be biased upwards. Basic summary statistics of flare characteristics are outlined in Table \ref{tab:summary_stats}.

In the subsequent analysis, we classify flare strength in a data-driven way, tied to the amount of spectral information content available and how well they constrain spectral fits (see Section \ref{sec:spectral_properties}; also see, e.g., \citetalias{Neilsen2013} and \citealt{Ponti2015} for alternative classification schemes). Through iterative testing, we consider strong flares to have fluence greater than 450 counts, moderate flares $100-450$ counts, and weak flares less than 100 counts.

\begin{deluxetable}{rccccc}
\tablewidth{0pt}
\tablecaption{Summary statistics of detected flare properties. $\mu$ is the mean and $\sigma$ is one standard deviation for the population. \label{tab:summary_stats}}
\tablehead{
\colhead{} & 
\colhead{$\mu$} & \colhead{$\sigma$} & \colhead{Median} & \colhead{Min} & \colhead{Max} 
}
\startdata
Duration (s) & 3035 & 2603 & 2224 & 367 & 14011\\
Max Rate (ct s$^{-1}$) & 0.12 & 0.22 & 0.05 & 0.02 & 1.88\\
Mean Rate (ct s$^{-1}$) & 0.06 & 0.08 & 0.03 & 0.01 & 0.69\\
Fluence (ct) & 184 & 425 & 70 & 16 & 3814\\
HR & 1.57 & 0.46 & 1.56 & 0.73 & 3.02
\enddata
\tablecomments{Weak flares have fluence less than 100 counts, moderate flares have fluence between 100 and 450 counts, and strong flares greater than 450 counts.}
\end{deluxetable}


\subsection{Flare Shape}
\label{sec:shape}
Flares distributed across the 25 years of \textit{Chandra} observations have similar morphologies. We characterize these shapes by the number of clearly distinguishable peaks in the flaring interval (single peaked, double peaked, complex substructure). The prevalence of rapid substructure in the sample, particularly in strong flares \citep[e.g.][]{Nowak2012, Haggard2019, Ghafourizadeh2026}, suggests that the X-ray emission closely tracks particle acceleration, favoring mechanisms with fast, local radiative response. Table \ref{tab:flare_complexity} shows that many, but not all, of the moderate flares have at least two peaks, and all strong flares have significant substructure. This brightness dependence may be an observational, rather than physical, effect however, as brighter flares have sufficient SNR to resolve temporal substructure.
Observational bias may therefore affect the number of visible peaks in weak flares. 
This is especially true for flares only slightly above the noise, which ``bubble'' with numerous short peaks (see 2008 July 26 light curve plot in Appendix \ref{appx:ex_lcs}). However, a general analysis of flare shapes may still reveal clues about horizon-scale features as we discuss in Section \ref{sec:morphological_similarities}.

{
\setlength{\tabcolsep}{12pt}
\begin{deluxetable}{lcccc}
\tablecaption{Flares are classified by their strength and shape. This table details the number of flares across each combination of classes. \label{tab:flare_complexity}}
\tablehead{
\colhead{} & 
\colhead{$N_T$} & \colhead{$N_S$} & \colhead{$N_D$} & \colhead{$N_{SS}$} 
}
\startdata
Weak & 62 & 50 & 12 & 0\\
Moderate & 30 & 7 & 23 & 0\\
Strong & 8 & 0 & 0 & 8
\enddata
\tablecomments{$N_T$ is the total number of flares in a strength class, $N_S$ is the number of single-peaked flares, $N_D$ is the number of double-peaked flares, and $N_{SS}$ is the number of flares with complex substructure beyond two peaks. Stronger flares appear more morphologically complex (though potentially as a result of SNR; see Section \ref{sec:shape}.)}
\end{deluxetable}
}



\section{Spectral Properties}
\label{sec:spectral_properties}
Moderate and strong flares (fluence $\gtrsim$ 100 counts) have sufficient signal for spectral analysis, where strong flares (fluence $\gtrsim 450$ counts) can independently constrain spectral parameters, whereas moderate flares with fluence between $100 - 450$ counts require combining their SNR with joint modeling. Joint fitting not only allows us to incorporate moderate flares to increase the sample size, but also enables us to probe how light curve properties such as luminosity may be connected to spectral characteristics. 

\subsection{Spectral Model}
\label{sec:spectral_model}
We use \texttt{XSPEC} and the spectral model from \citet{Haggard2019} to describe \Sg\ in the $2-10$~keV band as \textsc{pileup $\times$ tbabs $\times$ fgcdust $\times$ (powerlaw$_{f}$ + powerlaw$_{q}$ + bbody)}: a pileup component convolved with interstellar grain absorption (using \textsc{wlim} abundances), dust scattering, and the sum of a flaring power law, quiescent power law, and blackbody. The blackbody models contributions from the magnetar, and is set to zero when analyzing observations outside of the magnetar outburst. The quiescent power law is frozen at $\Gamma_q = 3$ as described in \citet{Nowak2012} and \citet{Haggard2019}. Its normalization is set to the quiescent level in the light curve divided by mean flare count rate, ${q}_{\text{quiescent}}/{q}_{\text{f,mean}}$. We also fix the hydrogen column density to 16.3$\times10^{22} \text{~cm}^{-2}$ \citep{Haggard2019}. We fit with the $\chi^2$ statistic.

For observations where the magnetar is in outburst, we independently fit the blackbody to the magnetar emission during periods when \Sg\ is quiescent in order to avoid biasing the magnetar fit. The best-fit magnetar temperature returned from this analysis is fixed in the subsequent \Sg\ spectral fit. 
We freeze the blackbody normalization in the \Sg\ spectral fit to the best-fit magnetar normalization scaled by the contamination level derived in Section \ref{subsec:magnetarcorrection}. Flare spectra and instrumental response files are generated with \textsc{specextract}.

\subsection{Independent Fitting of Strong Flares}
Eight flares have fluences exceeding 450 counts such that they can independently constrain parameters of the \Sg\ spectral model described above. Table \ref{tab:individual_flares} lists the results. All of the fits show agreement with the typical flaring power law index, $\Gamma_f = 2$ \citep{Nowak2012, Haggard2019}, within a 90\% confidence interval. 

\citet{Haggard2019} reported the same spectral parameters as listed Table \ref{tab:individual_flares} for flares in observation 15043 and 16218. Our $\Gamma_f$, $\alpha_{\text{pileup}}$, hardness ratio (HR), and $kT$ agree with their results, but our fluxes deviate slightly, likely due to our updated flux calibration in Section \ref{sec:flux_calibration}. We discuss the HRs further in Section \ref{sec:hr}. \citet{Nowak2012} analyzed flare 14392, and we reproduce their results within 90\% confidence intervals, including the fluxes. 


\begin{deluxetable*}{rccccccccccc}
\tablewidth{0pt}
\tablecaption{Spectral parameters of strong flares with SNR that can independently constrain the power law model outlined in Section \ref{sec:spectral_model}. \label{tab:individual_flares}}

\tablehead{
  \colhead{} &
  \multicolumn{4}{c}{\textbf{Flare Flux Properties}} &
  \multicolumn{4}{c}{\textbf{Flare Spectral Properties}} &
  \multicolumn{2}{c}{\textbf{ Magnetar}} & 
  \colhead{} \\
  \cline{2-5} \cline{7-9} \cline{11-11}
  \colhead{Flare ID} &
  \colhead{Fluence} &
  \colhead{$F_{\mathrm{2\text{--}8,abs}}$} &
  \colhead{$F_{\mathrm{2\text{--}8,unabs}}$} &
  \colhead{$F_{\mathrm{2\text{--}10,unabs}}$} &
  \colhead{ }&
  \colhead{$\alpha_{\mathrm{pileup}}$} &
  \colhead{$\Gamma_f$} &
  \colhead{HR} &
  \colhead{ } &
  \colhead{$kT$} &
  \colhead{$\chi^2$/DoF} \\
  \colhead{} &
  \colhead{(cts)} &
  \multicolumn{3}{c}{($\times~10^{-12}$ erg cm$^{-2}$ s$^{-1}$)} &
  \colhead{} &
  \colhead{} &
  \colhead{} &
  \colhead{} & 
  \colhead{} &
  \colhead{(keV)} &
  \colhead{}
}

\startdata
15043 & 3814 & $23.45^{+0.92}_{-0.90}$ & $68.82^{+2.67}_{-2.69}$ & $75.18^{+2.92}_{-2.93}$ &   & $0.67^{+0.15}_{-0.14}$ & $2.14^{+0.14}_{-0.14}$ & $1.68^{+0.06}_{-0.06}$ &   & $0.78^{+0.01}_{-0.01}$ & 146/149\\
20751 & 1381 & $7.48^{+0.40}_{-0.40}$ & $21.75^{+1.17}_{-1.17}$ & $24.96^{+1.34}_{-1.33}$ &   & 0.5 & $2.10^{+0.21}_{-0.21}$ & $1.52^{+0.08}_{-0.09}$ &   & -- & 61/68\\
14392 & 935 & $7.83^{+0.87}_{-0.82}$ & $23.89^{+2.64}_{-2.51}$ & $26.99^{+2.98}_{-2.83}$ &   & 0.5 & $2.25^{+0.37}_{-0.38}$ & $1.84^{+0.10}_{-0.11}$ &   & -- & 21/27\\
1561 & 924 & $2.70^{+0.22}_{-0.22}$ & $7.96^{+0.65}_{-0.64}$ & $9.13^{+0.74}_{-0.72}$ &   & 0.5 & $2.09^{+0.29}_{-0.29}$ & $1.86^{+0.12}_{-0.15}$ &   & -- & 46/44\\
16218 & 904 & $4.46^{+0.28}_{-0.28}$ & $13.70^{+0.86}_{-0.86}$ & $15.50^{+0.98}_{-0.97}$ &   & 0.5 & $2.23^{+0.24}_{-0.25}$ & $1.47^{+0.09}_{-0.11}$ &   & $0.77^{+0.02}_{-0.02}$ & 55/47\\
11843 & 559 & $15.81^{+2.00}_{-2.24}$ & $5.74^{+0.73}_{-0.81}$ & $7.95^{+1.00}_{-1.13}$ &   & $0.70^{+0.30}_{-0.38}$ & $1.87^{+0.45}_{-0.44}$ & $2.18^{+0.19}_{-0.20}$ &   & -- & 16/24\\
22595 & 465 & $2.11^{+0.16}_{-0.16}$ & $6.88^{+0.52}_{-0.51}$ & $7.62^{+0.57}_{-0.57}$ &   & 0.5 & $2.45^{+0.88}_{-0.82}$ & $1.23^{+0.11}_{-0.12}$ &   & -- & 3/5\\
13851 & 460 & $3.86^{+0.52}_{-0.50}$ & $11.35^{+1.53}_{-1.47}$ & $12.98^{+1.74}_{-1.68}$ &   & 0.5 & $2.12^{+0.54}_{-0.52}$ & $1.83^{+0.12}_{-0.16}$ &   & -- & 11/12\\
\enddata
\tablecomments{Each strong flare spectrum is fit with the model: \texttt{pileup $\times$ tbabs $\times$ fgcdust $\times$ (powerlaw$_{f}$ + powerlaw$_{q}$ + bbody)}, as described in Section \ref{sec:spectral_properties}. 
The flare in observation 14392 uses a grating, and we only fit the zeroth order photons (though its fluence includes both the first and zeroth orders). The grating and non grating observations have different flux calibrations, so lower fluence flares (observation 13851) may have higher fluxes than non gratings flares (e.g. observation 22595).}
\end{deluxetable*}

\subsection{Joint Fitting of Strong and Moderate Flares}
Joint fits apply the same model described in Section \ref{sec:spectral_model} but adopt a single flaring power law index for all observations, while retaining an independent normalization. Each spectrum is loaded in with its associated observation-dependent instrumental response file, which allows us to stack the flares to increase the SNR but still model observing configurations. We fit 37 of the strong and moderate flares together, and attempt a number of combinations based on various flare properties (for example, joint fitting of flares with mean flare rate greater than 0.1~ct s$^{-1}$). Notable results are summarized in Table \ref{tab:joint_fit}.

We provide two fits for all 37 flares: the first with $N_H$ fixed to 16.3$\times10^{22} \text{~cm}^{-2}$, and the second with it free. $N_H$ is degenerate with $\Gamma_f$ \citep{Yuan2018}, as shown in Table \ref{tab:joint_fit}. Our fit $N_H$ of $15.0\times10^{22} \text{~cm}^{-2}$ falls within the range found in other studies \citep{Nowak2012,Porquet2008,Bouffard2019}. However, we adopt a fixed $N_H = 16.3 \times 10^{22}$ cm$^{-2}$ in our analysis to facilitate direct comparison of $\Gamma_f$ with previous work.

\begin{figure}[]
\includegraphics[width=\columnwidth]{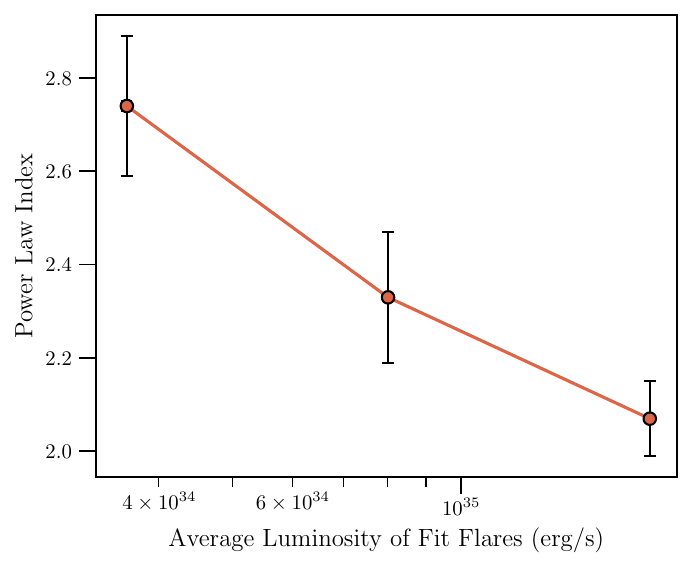}
\caption{The flaring power law indices of our joint spectral fits show an evolution from $\Gamma_f \sim 3$ to 2 as flares become more luminous. Each power law index is from the joint fit of multiple flares highlighted in the bottom three rows of Table \ref{tab:joint_fit}, and the luminosity of each data point is the average of the flares used in the spectral modeling. The uncertainties in average luminosity are too small to be seen.
\label{fig:pl_evolution}}
\end{figure}

Fitting by intervals of fluence, we measure a change in the flaring spectral index as flares get more luminous, from $\Gamma_f \sim 3$ to 2 (see Figure \ref{fig:pl_evolution} and Table \ref{tab:joint_fit}). \citet{Barriere2014} reported flares with spectral indices that differ beyond 90\% confidence ranges, and our analysis supports their result. The strength of this trend is more evident in our work compared to previous analysis because joint modeling increases the fitting sample size.


{
\setlength{\tabcolsep}{18pt}
\begin{deluxetable*}{rccccc}
\tablewidth{0pt}
\tablecaption{Joint spectral fits for various combinations of flares meeting the criteria listed in the first column. The first two rows report a fit of all moderate and strong flares (except observation 9173), while the three next rows correspond to subsets defined by fluence intervals.\label{tab:joint_fit}}
\tablehead{
\colhead{Flares Fitting Criteria} & \colhead{$\text{N}_{\text{flares}}$} & \colhead{$\alpha_{\text{pileup}}$} & \colhead{$n_{\rm H}$} & \colhead{$\Gamma_f$} & \colhead{$\chi^2$/DoF}}
\startdata
Fluence $>$ 100ct & 37 & 0.5 & 16.3 & $2.25^{+0.06}_{-0.06}$ & 1007/709 \\
Fluence $>$ 100ct & 37 & 0.5 & $15.0^{+1.0}_{-0.9}$ & $2.09^{+0.14}_{-0.14}$ & 1002/707 \\
\hline
Fluence $>$ 450ct & 8 & 0.5 & 16.3 & $2.07^{+0.08}_{-0.08}$ & 528/409\\
450ct $>$ Fluence $>$ 200ct & 9 & 0.5 & 16.3 & $2.33^{+0.14}_{-0.14}$ & 143/139\\
200ct $>$ Fluence $>$ 100ct & 20 & 0.5 & 16.3 & $2.74^{+0.15}_{-0.15}$ & 261/154\\
\enddata
\tablecomments{The flare in observation 9173 is excluded from joint fitting, despite having fluence $> 100$ counts, because its anomalously low mean count rate suggests that its duration may be mischaracterized.}
\end{deluxetable*}
}

\section{Discussion}
\label{sec:discussion}

The sample we collect in this \textit{Chandra} 25 year X-ray flare catalog enables population-level demographics that provide insight into the mechanisms underlying \Sg\ variability. The catalog contains 100 flares, 18 of which we report for the first time in this study (2 strong, 7 moderate, 9 weak), including detection of the second highest maximum count rate flare to date ($r_{max} = 0.89$ ct s$^{-1}$, L$_{2-10\text{keV}, \text{unabs}}$ $\sim$ 1.9$\times$10$^{35}$ erg s$^{-1}$; see Table \ref{tab:tablehead}). The other 82 have been previously reported in other works (\citealt{Nowak2012}, \citetalias{Neilsen2013}, \citealt{Ponti2015,Mossoux2017, Haggard2019, Mossoux2020}). 

\subsection{Systematics Influencing Reported Flare Properties}
\label{sec:flagging}

\begin{figure}
     \centering
     \includegraphics[width=\columnwidth]{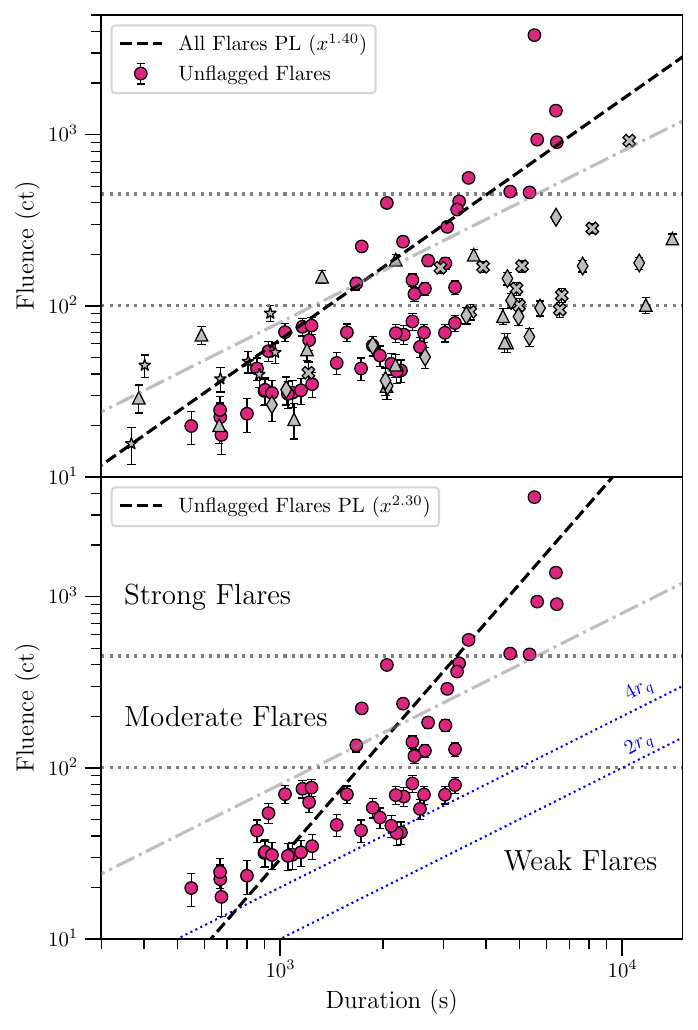}
    \caption{Correlation between the duration and fluence of a flare. 
    Black dashed lines are power law fits, and the grey dash-dotted line is a power law slope of $x^1$ to illustrate where the ratio of fluence to duration would be constant in log-log space. \textit{Top}: Flagged flares are identified by visual inspection, highlighted in grey. Triangle markers are flares cut off by the start or end of the observation. Stars are events where Bayesian Blocks has clearly underestimated the duration of the flare. Diamonds are faint flares where the specific start and end time is ambiguous because of low SNR. Cross markers represent flares reported as one, but potentially made up of two or more based on visual interpretation. \textit{Bottom}: The duration and fluence correlation of only unflagged flares. The blue dotted lines show where flares with mean count rates of 2 and 4 times the \Sg\ quiescent rate would lie to illustrate that our catalog contains a faint flare detection bias.}
    \label{fig:flare_cutting}
\end{figure}

Low SNR and observational effects such as observing window cutoffs may cause the Bayesian Blocks algorithm to inaccurately estimate the properties of a flare\footnote{We are confident in the presence of all flares reported in this study, but interpret the specific values of some flare properties with caution.}. We identify four main contributors to these discrepancies in our sample: (1) a flare may be cut off by the start or end of an observing window, 
(2) the exact start and end of a flare may be ambiguous due to low SNR, (3) the Bayesian Blocks algorithm may underestimate the duration of short flares with few data points, and (4) some events may be composed of at least two flares close in time, but are detected as one (see Appendix \ref{appx:ex_lcs} figures). We visually inspect the light curves of each detected flare for these systematics, and flag them (along with the reason why) if the discrepancy is present. Figure \ref{fig:flare_cutting} shows a summary of these flags.

All flares, regardless if they are flagged, are included in the demographic analysis portion of this work, but we highlight the flagged population to contextualize how they may impact the interpretation of our results. Though the classification of flag type is made strictly based on light curves, flagged events appear clustered in the duration-fluence parameter space of Figure \ref{fig:flare_cutting}. Specifically, flares with underestimated durations appear in the lower left of the top panel as some of the shortest flares of the population (star symbol in the top panel of Figure \ref{fig:flare_cutting}). Oppositely, the Bayesian Blocks algorithm may artificially increase the duration of low SNR, ambiguous flares because of uncertain start/end points relative to quiescence, where they appear as some of the longest events (Figure \ref{fig:flare_cutting} diamond).
The flares cut off by the observing window (Figure \ref{fig:flare_cutting} triangles), and events potentially composed of at least two flares (Figure \ref{fig:flare_cutting} crosses) are also distributed as we expect i.e., randomly located across parameter space, because their underlying causes are not related to SNR or flare duration. The clustering of these flagged flares indicates that observational effects may be influencing the observed spread of flare properties, and our flagging is therefore intended as a way to engage with selection effects and detection biases.

\subsection{Parameter Distributions}
\label{sec:parameter_distributions}
\begin{figure}
     \centering
     \includegraphics[width=\columnwidth]{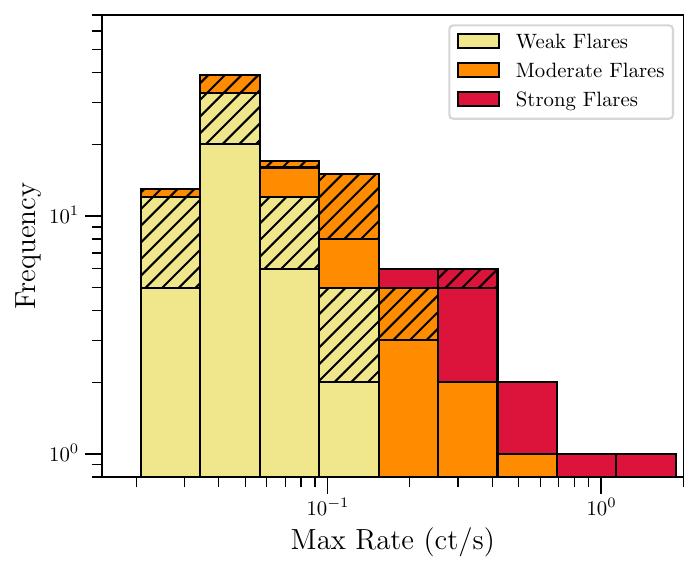}
     \includegraphics[width=\columnwidth]{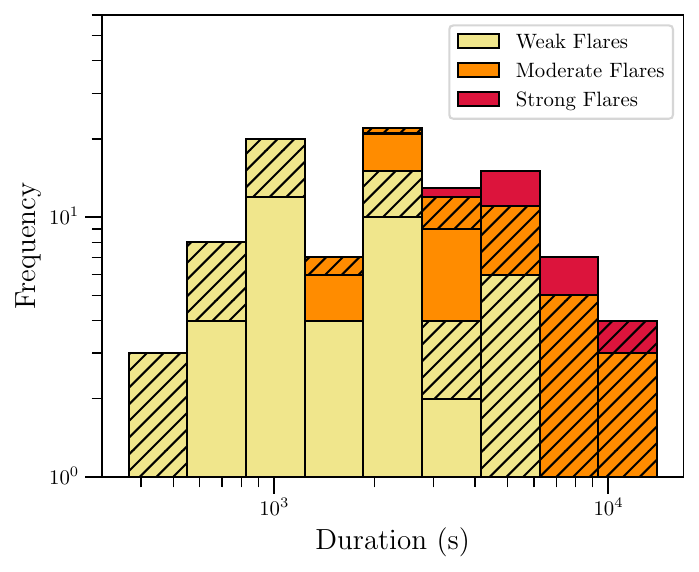}
    \caption{Distributions of maximum count rate and duration. Hashed sections represent flagged flares, as described in Section \ref{sec:flagging}. \textit{Top}: The distribution of maximum count rates shows that bright flares are significantly less common than weak ones. \textit{Bottom}: Unflagged flares appear to be 100s to 1000s of seconds long, and strong flares have longer durations. Most flagged flares are long in duration, likely because Bayesian Blocks overestimates the duration when it is less clear where the change points are.}
    \label{fig:histograms}
\end{figure}

Figure \ref{fig:histograms} shows that weak flares are more common than moderate and strong flares in our catalog (62\%, 30\%, and 8\%, respectively). The bottom panel of also highlights that strong flares tend to be longer in duration than weaker flares. 
\citet{Neilsen2015} estimated that $\sim 10 - 15\%$ of flares may be too faint to be detected, and blend with the quiescent level, so the top panel of Figure \ref{fig:histograms} may in reality have a larger number of low rate flares than what we detected here. Both of these results are consistent with what was reported by \citetalias{Neilsen2013}, where in this work, we have enriched \citetalias{Neilsen2013}'s distributions with our additional statistics.

\subsection{Hardness Ratios}
\label{sec:hr}
\begin{figure}[]
\includegraphics[width=\columnwidth]{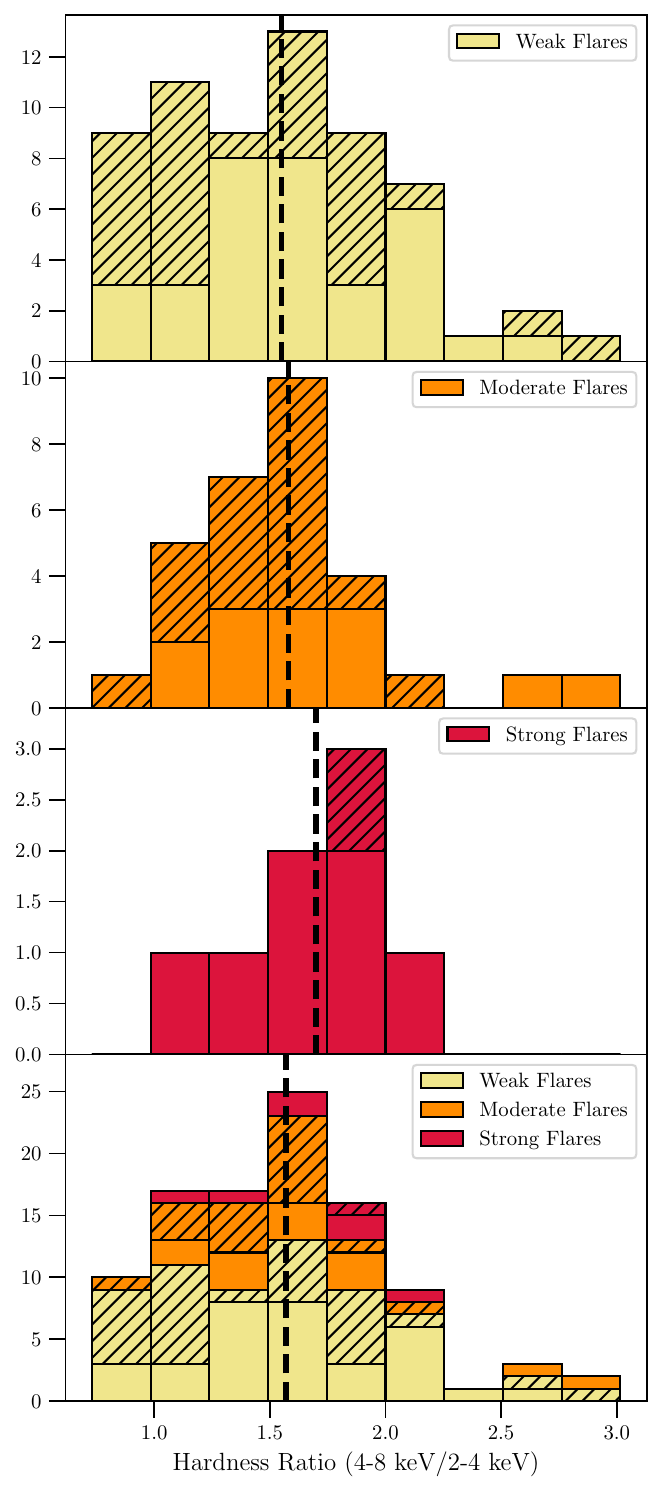}
\caption{Distributions of the 4--8~keV/2--4~keV hardness ratio (HR) for weak (top), moderate (top middle), strong (bottom middle), and all (bottom) flares. Hashed sections represent flagged flares, as described in Section~\ref{sec:flagging}. The HRs are roughly normally distributed, with slightly larger mean HR for strong flares than weak ones (black dashed lines). This indicates that the proportion of high energy photons changes between strength classes. 
\label{fig:hr}}
\end{figure}

We extract the number of $4-8$~keV and $2-4$~keV photons in each pileup corrected-flare, then use the \texttt{BEHR} code \citep{Park2006} to estimate their $4-8$ keV/$2-4$ keV hardness ratios (HRs) and associated uncertainties. Figure \ref{fig:hr} shows these results grouped by flare strength. 
There are few strong flares, which limits the robustness of HR statistics for that class, and the weak flares suffer from low rate detection bias \citep{Neilsen2015}. We find that the average quiescent state HR is $\sim 0.74 \pm 0.01$, much lower than the average HR of all flares, HR = $1.57^{+0.22}_{-0.38}$, from Table \ref{tab:summary_stats}. It is therefore likely that the HR of additional faint flares, with mean count rate less than about 4 times the quiescent rate (see the bottom panel of Figure \ref{fig:flare_cutting}), would populate the left side of the top panel in Figure \ref{fig:hr}. Based on this, the distribution of HRs appears to be somewhat Gaussian for all flare strengths. 
We find the mean HR = $1.55^{+0.03}_{-0.07}$ (weak flares), $1.58^{+0.04}_{-0.05}$ (moderate flares), and $1.70^{+0.04}_{-0.05}$ (strong flares). While strong flares appear somewhat harder on average (consistent with the findings in \citetalias{Neilsen2013}), the differences are modest and potentially suggest a continuous trend of spectral hardening with flare strength rather than distinct spectral regimes. 
\cite{Nowak2012} and \cite{Haggard2019} also computed HRs for individual flares. Our calculations agree with their results (see Table \ref{tab:tablehead}), but they do not report enough flares to study population-level HR trends.

The modest HR dependence on flare strength is supported by the spectral slope evolution shown in Figure \ref{fig:pl_evolution}. Changing the spectral index between flare types implies that weak flares contain fewer high energy photons than strong flares, ($\Gamma_{f, \text{weak flare}} > \Gamma_{f, \text{strong flare}}$), which is reflected in Figure~\ref{fig:hr} by the lower average HR for the weak flare distribution. The measured HRs validate the inferred spectral hardening with increasing flare strength.



\subsection{Parameter Correlations}\label{subsec:corr}

\begin{deluxetable*}{rccccc}
\tablewidth{0pt}
\tablecaption{Pearson coefficient, $\rho$, for correlations among flare properties. \label{tab:correlations}}
\tablehead{
\colhead{Correlation} & 
\colhead{All Flares} & \colhead{Unflagged Flares} & \colhead{\citetalias{Neilsen2013}} 
}
\startdata
Duration/Max rate & 0.24 & 0.65 & 0.27\\
Duration/Fluence & 0.32 & 0.65 & 0.54\\
Fluence/Max Rate & 0.97 & 0.98 & 0.89
\enddata
\end{deluxetable*}

\begin{figure}
     \centering
     \includegraphics[width=\columnwidth]{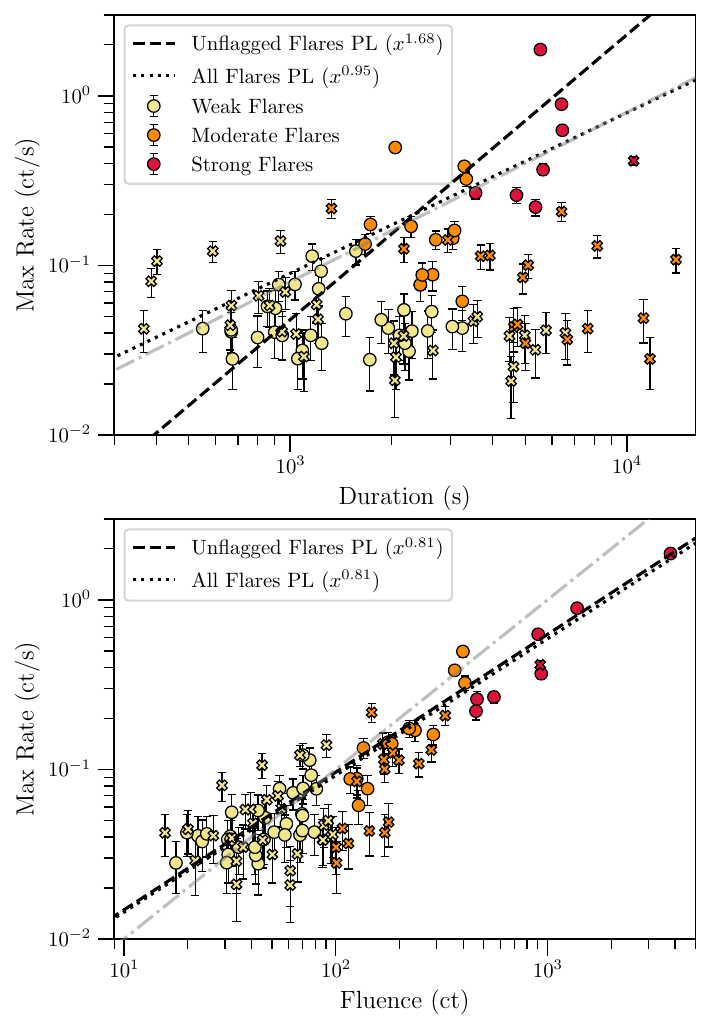}
    \caption{Correlations between duration, fluence, and maximum count rate. Events marked with a cross are flagged flares from Figure \ref{fig:flare_cutting}. \textit{Top}: The duration and maximum count rate of flares is moderately correlated. Many flagged flares appear at long duration and low count rate, as discussed in Section \ref{sec:flagging}. A power law fit to all flares is given by the dotted line, and to only unflagged flares by the dashed line. The grey dash-dotted line is $x^1$. \textit{Bottom}: There is a strong correlation between fluence and maximum count rate. The dashed, dotted and dash-dotted lines are the same as the top panel. Many events in the lower left region with higher spread are flagged, potentially indicating that they deviate from the power law fit due to systematics instead of inherent flare properties.
    }
    \label{fig:param_space}
\end{figure}

We report the Pearson correlation coefficient ($\rho$) for the cases considering all flares and only unflagged flares, as well as compare them with \citetalias{Neilsen2013} in Table \ref{tab:correlations}. Excluding flagged flares increases the strength of the correlation in all relationships. 
Figures \ref{fig:flare_cutting} and  \ref{fig:param_space} show the correlated parameters, and we fit a power law to investigate how the properties of flares evolves as a function of strength. In the duration–fluence relation (Figure \ref{fig:flare_cutting}), we measure a power-law slope of $1.4 \pm 0.01$ for all flares, which increases to $2.3 \pm 0.02$ when considering only unflagged events. Similarly, we find slopes of 0.95 $\pm$ 0.02 and 1.68 $\pm$ 0.03 for the correlation in duration and maximum rate for all and unflagged flares, respectively. We observe a consistent power law slope of 0.81 $\pm$ 0.01 for all and unflagged flares in the fluence and maximum rate correlation. The power-law slope increases when flagged flares are omitted for correlations involving duration, likely because many of the removed events are faint and long. Slopes greater than unity in the duration–fluence and duration–maximum rate relations indicate that longer flares are intrinsically more energetic and reach higher peak rates, rather than simply being time-extended versions of weaker events that would follow a power law slope of 1. 

Strong, unflagged flares exhibit a characteristic timescale of $\sim5$~ksec, comparable to the findings in \citetalias{Neilsen2013} (see the top panel in Figure \ref{fig:param_space}). Most flares that exceed this timescale are flagged. This does not entirely rule out the possibility of flares longer than 5~ksec, as the flagging only identifies skepticism in the properties of the event. Instead, it adds credibility to the previously reported timescale. 

The top left region of Figure \ref{fig:param_space} is mostly empty, indicating a scarcity of short, strong flares. This may be a natural consequence of energy injection limits when creating flares. \citet{vonFellenberg2023} reported a characteristic shape of \Sg\ flares from \textit{Chandra} that agrees with this sentiment, where flares ramp up to a maximum over a similar timescale. Short bursts to extreme count rates are not observed.

\subsection{Morphological Similarities}
\label{sec:morphological_similarities}
\begin{figure}
     \centering
     \includegraphics[width=\columnwidth]{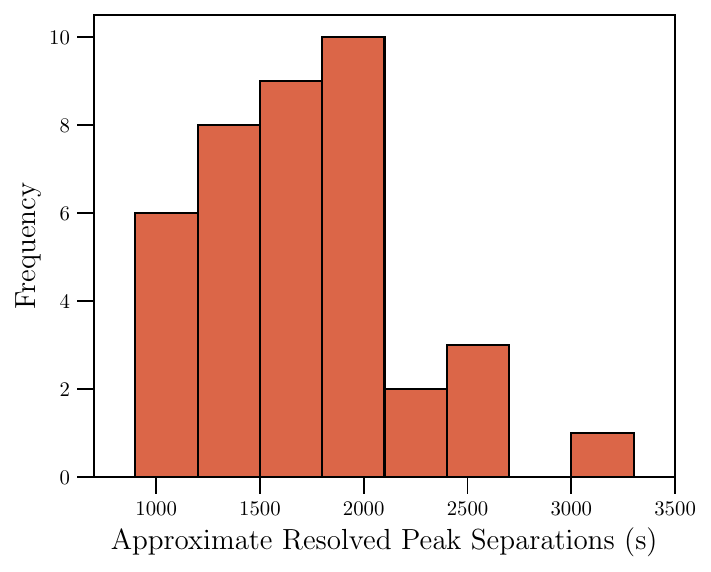}
     \caption{Separation time between the two most significant peaks in double and higher substructure flares (when they can be confidently resolved). We visually inspect the flare light curve for the maxima of two peaks, and count the number of bins between them. We observe that these peaks are often separated by about $\sim$1$-$2 ks, even though the flares are independent.}
     \label{fig:sephist}
\end{figure}

Some double-peaked flares identified in Section~\ref{sec:shape} appear as two different shapes: ``shoulders'' (see the 2019 August 19 light curve in Appendix \ref{appx:ex_lcs}) and equal double-peaks (2013 September 14 and 2013 October 28 light curves; Appendix \ref{appx:ex_lcs}). The shoulder structures are secondary bumps on the leading or falling side of the main flare, with no apparent preference for which side the shoulder is on. In equal double-peaked flares, where the primary and secondary bump are more similar in strength, we observe that the separation timescale between these peaks clusters around $\sim$1$-$2 ks (see Figure \ref{fig:sephist}). This double-peaked structure is independently-confirmed by other instruments (e.g. XMM-Newton observations of Sgr A* flares in \citet{Ghafourizadeh2026}), suggesting the structural morphology is intrinsic rather than an instrumental effect. 

The consistent shape/temporal spacing of these double-peaked flares may derive from orbital motion of a hotspot \citep{Hamaus2009, Karssen2017, GRAVITY2018, GRAVITY2020, GRAVITY2021, GRAVITY2023, vonFellenberg2023, vonFellenberg2024, Michail2024}. \cite{Karssen2017} found that in double-peak events, one maximum is created by lensing when the hotspot is behind the black hole, and the other by Doppler boosting as it approaches the observer in its orbit. They find that at a fixed inclination, the height and width of each sub-peak will depend on the orbital distance and size of the hotspot (see Figure 2 in \citealt{Karssen2017}). However, similar multi-peaked structures may also arise from intermittent energy release in magnetic reconnection or turbulent variability. It may be possible to analyze a population of emitting-region properties through double-peaked X-ray flares. The precise modeling of these parameters is outside the scope of this work, but the addition of the new, bright multi-peaked flares we present in this study motivates this future analysis.

\subsection{Spectral Differences}
The combined evidence of spectral slope evolution (Section~\ref{sec:spectral_properties}), the dependence of hardness ratio on flare strength (Section~\ref{sec:hr}), and the duration-fluence and duration-maximum rate correlations (Section~\ref{subsec:corr}) indicate that strong flares are not just longer versions of weak flares (e.g., constant count rate-to-duration ratio). Instead, stronger flares are systematically harder; producing a greater fraction of high-energy photons per unit duration. This establishes a coupling between spectral properties and flare energetics that theoretical models must reproduce. X-ray flares are diagnostically powerful, as the highest-energy emission in the brightest events is where degeneracies among acceleration and radiative models begin to break down \citep{Dodds-Eden2009, Ponti2017}. Observationally, flare spectra are known to vary both between events and within individual flares \citep{Barriere2014}, yet modeling efforts have largely focused on individual events, either by testing whether specific acceleration mechanisms (e.g., magnetic reconnection, turbulence, or shocks) can produce the required nonthermal particle distributions, or by fitting flare spectra with assumed radiative process (typically synchrotron or synchrotron self-Compton). In many cases, different combinations of these assumptions can reproduce the properties of a single flare \citep{Eckart2012, GRAVITY2021, Boyce2022}. 


Population-level studies provide a more constraining test by calling models to reproduce the statistical properties of the flare population. \citet{Dibi2016} demonstrated how population statistics might be used to test radiative scenarios, by comparing observed near-IR and X-ray flux distributions to predictions from synchrotron self-Compton and synchrotron models. The dataset presented here enables this framework to be extended to incorporate coupled spectral, temporal, and energetic properties of the X-ray flare population. 
At the same time, kinetic studies have shown that particle spectra are sensitive to plasma conditions, such as guide field strength (S. S\'{a}nchez-Maes et al., in prep.), and the balance between turbulent heating and radiative cooling \citep{Uzdensky2018} can produce a range of particle spectral slopes. Global simulations incorporating nonthermal particle prescriptions further show that predicted spectra depend on the assumed energy injection model \citep{Chatterjee2021}. In this context, the trends identified here provide observational constraints on how these physical variations manifest across the flare population.

\section{Conclusion}
\label{sec:conclusion}

In this work, we systematically reanalyze 6.8~Ms of \Sg\ observations with the \textit{Chandra} X-ray Observatory and detect 100 flares. 18 are reported for the first time in this study. We develop a processing and characterization pipeline incorporating updated calibration, treatment of contamination from the Galactic Center magnetar, and instrumental corrections, including a light curve pileup estimation technique based on the \textit{Chandra} ray-tracing simulator \texttt{MARX}.  This work presents the most complete \textit{Chandra} catalog of \Sg\ X-ray flares to date, expanding on previous surveys \citet{Mossoux2017} and \citet{Mossoux2020}.

Using this sample, we conduct a population level analysis of flare properties that reveals: 

\begin{enumerate}
    \item The spectral index of flares transitions from $\Gamma_f \sim 3$ to $\sim 2$ as flares get brighter.
    \item Hardness ratios appear roughly normally distributed, with stronger flares modestly harder on average.
    \item Flare morphology becomes more complex with increasing strength, as well as two double-peaked morphologies being consistently observed across the 25 year dataset.
    \item Correlations in duration-maximum count rate, duration-fluence, and fluence-maximum count rate that were previously reported by \citetalias{Neilsen2013} are reinforced with improved statistics.
\end{enumerate}
In the context of X-ray flare emission, 
these four phenomena provide a population-level constraint that models must satisfy, extending beyond fits to individual flares. 

The analysis presented in this work acts as a comprehensive study of 25 years of dedicated \textit{Chandra} observations of \Sg. As multi-wavelength campaigns continue to evolve, we encourage further investigation into the temporal correlations and connections across wavelength regimes, and are optimistic this systematic demographic study of X-ray flares will serve as a reference for understanding how \Sg\ variability manifests across physical scales.

\begin{acknowledgments}

We would like to give a special thank you to Melania Nynka and Moritz G\"{u}nther for their informative discussions about \textit{Chandra} and pileup.

We also acknowledge the dedicated efforts of Chloé Robeyns, Elisa Jacquet, Jasmine Zhang, and Maude Lariviére for their contributions to the \textit{Chandra} \Sg\ X-ray flare analysis pipeline, which help lay the foundation for the data processing presented in this study.

This research was supported by the International Space Science Institute (ISSI) in Bern, through ISSI International Team project \#24-610, and we thank the ISSI team for their generous hospitality.

This research made use of data obtained from the Chandra Source Catalog, provided by the Chandra X-ray Center (CXC). 

DH, ZS, NMF, MB, SDvF acknowledge support from the Canadian Space Agency (23JWGO2A01 and 25JWGO4A01), the Natural Sciences and Engineering Research Council of Canada (NSERC) Discovery Grant program, the Canada Research Chairs (CRC) program, the Fondes de Recherche Nature et Technologies (FRQNT) Centre de recherche en astrophysique du Québec, and the Trottier Space Institute at McGill.
ZS is supported by the Chalk-Rowles fellowship. NMF acknowledges funding from the FRQNT Doctoral Research Scholarship and NSERC Canada Graduate Research Scholarship.
JMM is supported by an NSF Astronomy and Astrophysics Postdoctoral Fellowship under award AST-2401752. 
SDvF gratefully acknowledges the support of the Alexander von Humboldt Foundation through a Feodor Lynen Fellowship and thanks CITA for their hospitality and collaboration.
SDvF acknowledge the support of the Natural Sciences and Engineering Research Council of Canada (NSERC), [funding reference number 568580] Cette recherche a \'et\'e financ\'ee par le Conseil de recherches en sciences naturelles et en g\'enie du Canada (CRSNG), [num\'ero de r\'ef\'erence 568580].
\end{acknowledgments}

\facilities{CXO}

\software{Astropy\citep{astropy:2013, astropy:2018, astropy:2022}, BEHR \citep{Park2006}, CIAO \citep{Fruscione2006}, MARX \citep{Davis2012}, XSpec \citep{Arnaud1999}, WebPIMMS}

\appendix 

\section{Pileup Correction}\label{appx:pileup}

\begin{figure*}[htb]
     \centering
     \includegraphics[width=\textwidth]{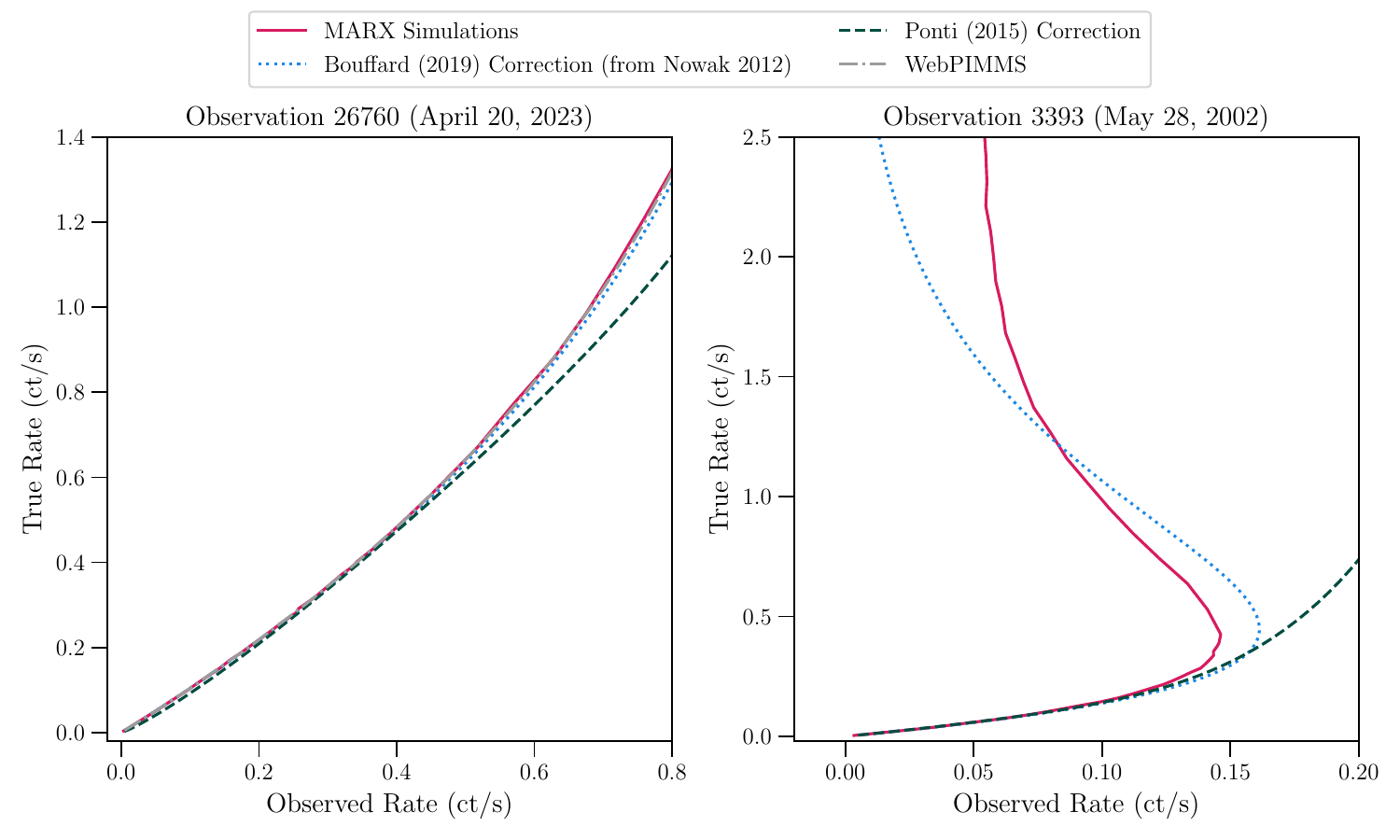}
     \caption{Pileup corrections for two example observations. We use \texttt{MARX} to calculate a true flux (pileup corrected) for every observed flux. \textit{Left}: Observation 26760 was captured 24 years into \textit{Chandra}'s lifetime, on the ACIS-S instrument with no grating and a 0.4~s/frame exposure time. Its pileup is relatively minor, and our correction (solid pink) agrees with treatments proposed by \citet{Bouffard2019}, \citet{Nowak2012}, and \citet{Ponti2015}. (\texttt{WebPIMMS} is an online tool to convert \textit{Chandra} count rates to physical units, and correct for pileup\footnote{https://cxc.harvard.edu/toolkit/pimms.jsp}). \textit{Right}: Observation 3393 was captured on 2002 May 28, with the ACIS-I instrument, no grating, and a 3.1~s/frame exposure time. It exhibits burn-out turnover, where large true fluxes appear as low observed fluxes because the detector incorrectly discards piled photons that it believes are errors. Burn-out can be visually identified in images as well. \citet{Bouffard2019} and \citet{Ponti2015} do not have \textit{Chandra} mission lifetime dependent corrections, and therefore do not always agree with our pileup corrections.}
     \label{fig:marx}
\end{figure*}

Section \ref{sec:pileup} describes our pileup correction technique, where we use the \texttt{MARX} \textit{Chandra} ray-tracing simulator to estimate how each observation responds to pileup given their unique observing conditions. Figure \ref{fig:marx} shows the pileup response for two example observations. We find that the frame time, instrument, presence of the HETG, and time in \textit{Chandra's} mission lifetime control the response. In heavily piled regimes, the detector can ``burn-out'', where a void forms at the center of the bright source because of grade migration and energy detection limits, decreasing its count rate. This phenomenon is discussed in the \textit{ABC Guide to Pileup}\footnote{https://cxc.harvard.edu/ciao/download/doc/pileup\_abc.pdf}. We observe this effect in \texttt{MARX}, with the count rate turning over as the flux gets sufficiently strong (see the right panel of Figure \ref{fig:marx}).

\section{Example light curves}\label{appx:ex_lcs}
\begin{sidewaysfigure}
\centering
\includegraphics[width=\textheight]{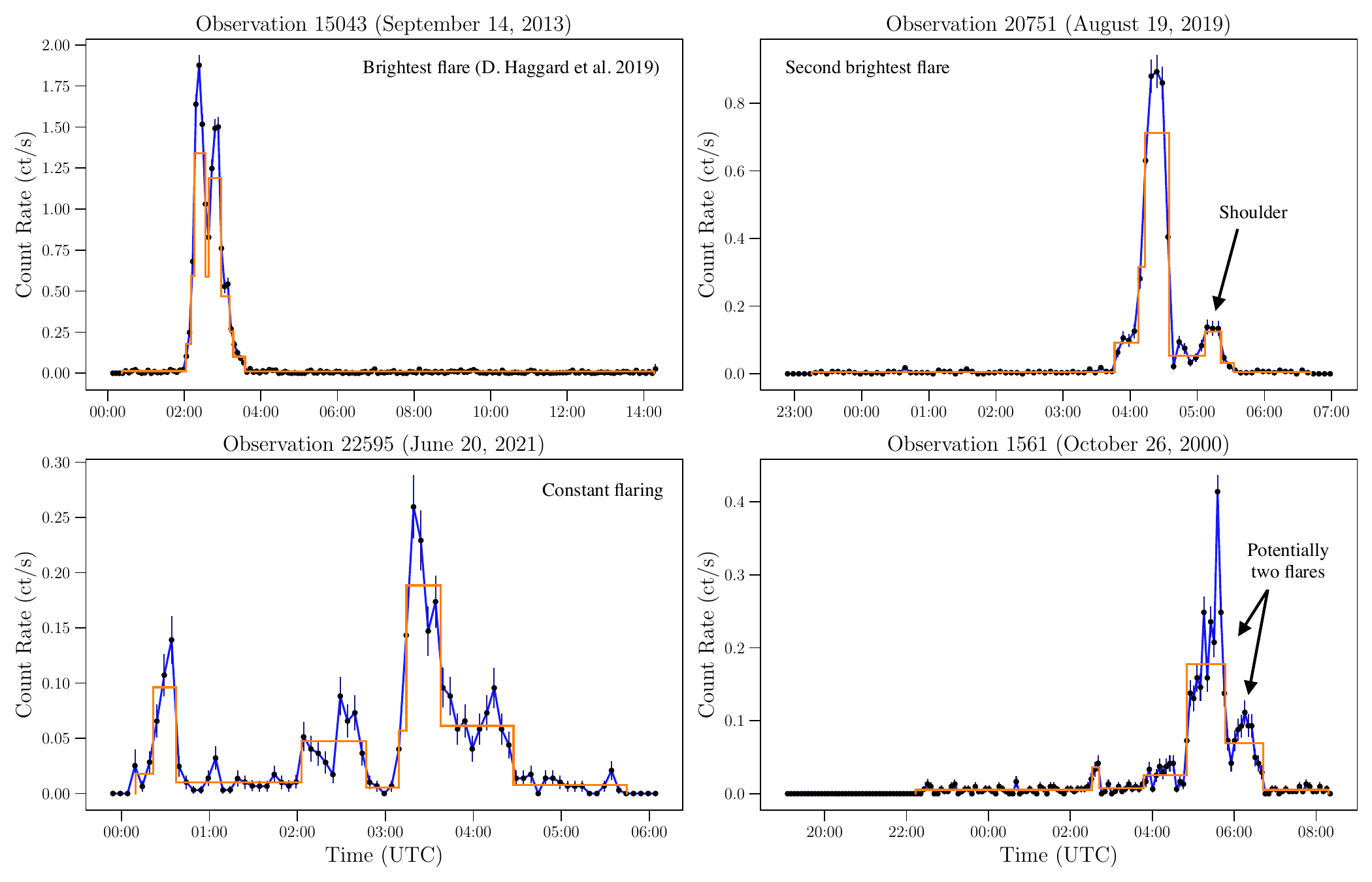}

\caption{Four example \Sg\ light curves (blue) overlaid with Bayesian Blocks (orange). \textit{Top left}: Observation 15043 contains the brightest detected \textit{Chandra} flare and was analyzed in \citet{Haggard2019}. \textit{Top right}: We present for the first time the second strongest \textit{Chandra} flare captured in observation 20751 on 2019 August 19. It also shows the shoulder structure we introduce in Section \ref{sec:morphological_similarities}, with one strong peak and a weaker secondary bump on one side. \textit{Bottom left}: Some observations, such as 22595, taken on 2021 June 20, contain variability for almost the entire observing period. This makes the quiescent rate difficult to determine. \textit{Bottom right}: We flag flares that were reported as one, but may actually be two or more (see Figure \ref{fig:flare_cutting}). Observation 1561, captured on 2000 October 26, is $\sim 2$ times longer than typical strong flares, and contains multiple resolved peaks, and was flagged as a multi-flare event.}
\label{fig:light curves1}
\end{sidewaysfigure}

\begin{sidewaysfigure}
\centering
\includegraphics[width=\textheight]{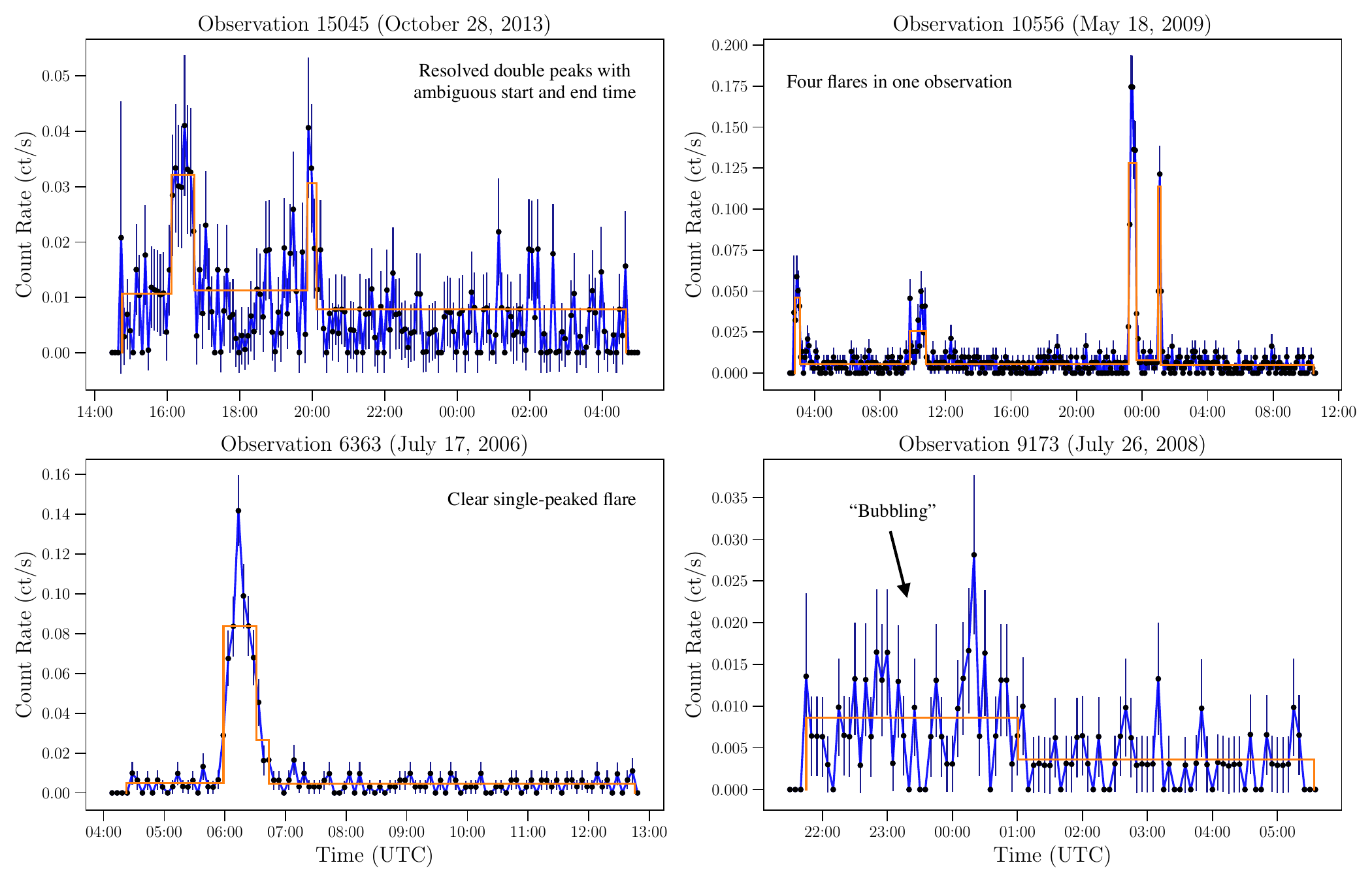}
\caption{Four additional example light curves (blue) with Bayesian Blocks (orange). \textit{Top left}: The flare in observation 15045 is an example of the case where the specific start and end times are ambiguous. \textit{Top right}: Some flares, such as Observation 10556, which was taken on 2009 May 18, contain more than one independent flare. \textit{Bottom left}: The shape of flares varies, especially across SNR. Observation 6363 is an example of a resolved single peak event. \textit{Bottom right}: Observation 9173, captured on 2008 July 26, is an additional example of a flare with an ambiguous start and stop time. This flare shows the ``bubbling'' phenomenon, where it has low-level activity above quiescence, but no clear peaked structure. The flare flagging we perform in this study (see Figure \ref{fig:flare_cutting} and Section \ref{sec:flagging}) attempts to highlight that the properties of flares like this are difficult to determine.}
\label{fig:light curves2}
\end{sidewaysfigure}

We generate light curves for every observation, and visually inspect them to understand observational trends for considerations such as flagging in Figure \ref{fig:flare_cutting}. We present 8 example light curves in Figures \ref{fig:light curves1} and \ref{fig:light curves2}.

\clearpage


\bibliography{sample701}{}
\bibliographystyle{aasjournalv7}

\section{Observation Parameters}
Test
\begin{deluxetable*}{rllllll}
\tablewidth{0pt}
\tablecaption{The observing setups for archival \Sg\ observations. All modes are in exposure mode TE and data mode FAINT. Tracking these modes is necessary to understand where certain corrections are necessary, such as grating regions. \label{tab:observingmodes}}
\tablehead{
\colhead{Observing Mode} & \colhead{Instrument} & \colhead{Grating} & \colhead{CCDs} & \colhead{Frame Time (s)} & \colhead{Subarray Type} 
}
\startdata
TE\_001B4 & ACIS-I & None & I0, I1, I2, I3, S2, S3 & 3.2 & None\\
TE\_002DC & ACIS-I & None & I0, I1, I2, I3, S2 & 3.2 & None\\
TE\_002FA & ACIS-I & None & I0, I1, I2, I3 & 3.1 & None\\
TE\_003B6 & ACIS-I & None & I0, I1, I2, I3, S2 & 3.1 & None\\
TE\_007F4 & ACIS-I & None & I0, I1, I2, I3 & 3.2 & None\\
TE\_00866 & ACIS-I & None & I0, I1, I2, I3 & 3.1 & None\\
TE\_008D0 & ACIS-S & HETG & S1, S2, S3, S4, S5 & 3.1 & None\\
TE\_008FC & ACIS-S & None & S3 & 0.4 & 1/8 \\
TE\_00C2C & ACIS-S & None & S3 & 0.6 & 1/8 \\
TE\_00D50 & ACIS-S & None & S2, S3 & 0.5 & 1/8 \\
\enddata
\tablecomments{Observing modes TE\_002FA and TE\_00866 differ in the ``Spectra Max Count'' - a limit on the number of events per frame per spectrum.}
\end{deluxetable*}
\newpage



\end{document}